\documentclass[aps,prd,twocolumn,english,nofootinbib]{revtex4-1}
\usepackage[T1]{fontenc}
\usepackage[latin9]{inputenc}
\usepackage{natbib}
\usepackage{color}
\usepackage{url}
\usepackage{babel}
\usepackage{array}
\usepackage{float}
\usepackage{amsmath}
\usepackage{ulem}
\usepackage{stmaryrd}
\usepackage{stackrel}
\usepackage{graphicx}
\usepackage{tikz}
\usepackage{multirow}
\usepackage{comment}
\usepackage[toc,page]{appendix}

\newcommand{\vect}[1]{\boldsymbol{#1}}
\usetikzlibrary{shapes.geometric, arrows}

\tikzstyle{startstop} = [cylinder, rounded corners, minimum width=1.5cm, minimum height=0.5cm, text width=0.8cm, text centered, draw=black, fill=blue!30, shape border rotate=90]
\tikzstyle{io} = [trapezium, trapezium left angle=70, trapezium right angle=110, minimum width=4cm, minimum height=1cm, text width=2.5cm, text centered, draw=black, fill=blue!30]
\tikzstyle{process} = [rectangle, text width=3cm, minimum width=3cm, minimum height=1cm, text centered, draw=black, fill=yellow!30]
\tikzstyle{processl} = [rectangle, text width=3cm, minimum width=3cm, minimum height=1cm, text centered, draw=black, fill=green!30]
\tikzstyle{postprocess} = [rectangle, text width=3.8cm, minimum width=4cm, minimum height=1cm, text centered, draw=black, fill=yellow!30]
\tikzstyle{postprocessl} = [rectangle, text width=3.8cm, minimum width=4cm, minimum height=1cm, text centered, draw=black, fill=green!30]
\tikzstyle{exprocess} = [rectangle, text width=3cm, minimum width=3cm, minimum height=1cm, text centered, draw=black, fill=red!30]
\tikzstyle{exprocessl} = [rectangle, text width=3.8cm, minimum width=4cm, minimum height=1cm, text centered, draw=black, fill=red!30]
\tikzstyle{inprocess} = [rectangle, text width=2.5cm, minimum width=3cm, minimum height=1cm, text centered, draw=black, fill=green!30]
\tikzstyle{decision} = [diamond, minimum width=1.5cm, minimum height=0.7cm, text width=2.08cm, text centered, draw=black, fill=green!30]
\tikzstyle{arrow} = [thick, ->, >=stealth]
\tikzstyle{arrowboth} = [thick, <->, >=stealth]

\usepackage[unicode=true,pdfusetitle,
 bookmarks=true,bookmarksnumbered=false,bookmarksopen=false,
 breaklinks=false,pdfborder={0 0 1},backref=false,colorlinks=true]{hyperref}
\hypersetup{
 linkcolor=blue,urlcolor=blue,citecolor=blue}
\makeatletter

\usepackage{blindtext}

\makeatother
\begin{document}
\title{Observable spins in gravitational waves from compact binary mergers}
\author{Souradeep Pal}
\email{sp19rs015@iiserkol.ac.in}

\affiliation{Indian Institute of Science Education and Research Kolkata, Mohanpur,
Nadia - 741246, West Bengal, India}
\begin{abstract}
We investigate the measurability of effective inspiral spin in the detectable compact binary mergers using gravitational-wave observations. Measurements from the latest gravitational-wave transient catalog do not rule out the existence of binary systems with non-zero effective spins. However, we observe an apparent correlation between the inferred effective inspiral spin and the loudness of the gravitational-wave events-- loud events typically have close-to-zero effective spins whereas fainter events tend to be inferred with relatively arbitrary effective spins. Through simulations, we demonstrate that non-negligible effective spins can be systematically inferred from non-spinning systems at small signal strengths. These two observations support the possibility that the effective spin magnitudes in the observable compact binaries are generally small. Future detections can have potential impact on the understanding of their population and other astrophysical inferences.
\end{abstract}
\date{\today}
\maketitle
\section{Introduction}\label{sec:introduction}
Observation of gravitational-wave (GW) signals from compact binary mergers has become a routine effort. Component spins of the binary systems can play a crucial role in the signal detection, the estimation of the source properties, and can potentially inform us about their origin.

The first direct detection of GW signal from a compact binary merger was made in 2015 by the Laser Interferometer Gravitational-wave Observatories (LIGOs)~\citep{abbott2016observation}. The fourth observing run (O4) of the Advanced LIGO, Advanced Virgo, and KAGRA began on May 24, 2023. A catalog of gravitational-wave transient (GWTC) events until the first part of the run (O4a) has been released~\citep{ligo2025gwtc}. It contains more than 200 confident candidates of astrophysical events and their properties. Several exceptional events are also reported~\citep{kw5gd732,Abac_2025}. A majority of these events are consistent with signals from the merger of astrophysical black holes (BHs)~\citep{ligo2025gwtc}. Several others are consistent with neutron star (NS) mergers~\citep{abbott2017gw170817,Abbott_2020,abbott2021observation,ligo2025gwtc}. Overall, a few very loud events have been reported.

\begin{figure}[t]
\begin{centering}
\includegraphics[scale=0.68]{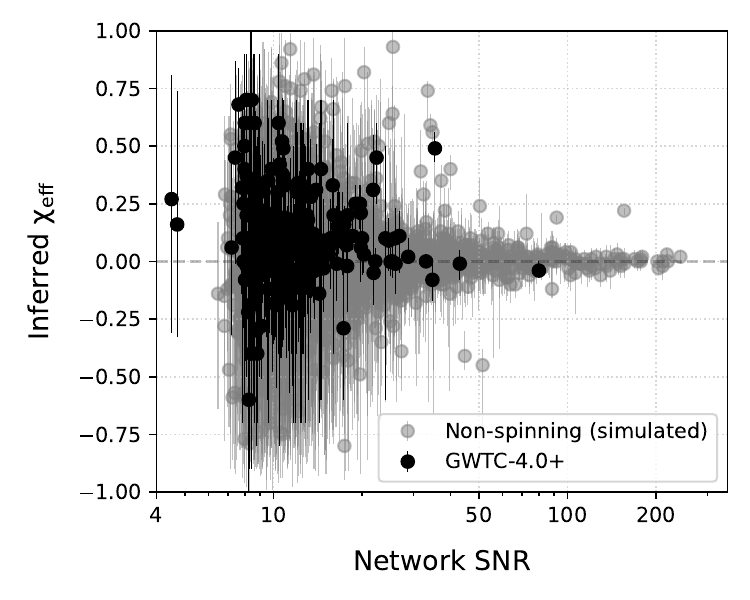}
\par\end{centering}
\vspace{-0.21cm}
\caption{Distribution of the inferred effective inspiral spin ($\chi\textsubscript{eff}$) with the network SNR for simulated non-spinning merger events (in grey). The same is shown for the confidently detected GW events (in black), which contains events reported in \text{GWTC-4.0}, as exceptional and other events~\citep{gwoscevents}, collectively referred to as \text{GWTC-4.0+}. A circle indicates the inferred value and the error bars represent the 90\% credible interval of the measurement. Essentially, note that non-negligible $\chi\textsubscript{eff}$ values can be inferred from non-spinning sources at generally small SNRs, as discussed later in the text.}
\vspace{-0.15cm}
\label{fig:plotcomb}
\end{figure}

For a binary system, with component masses $\mathrm{m}\textsubscript{1}$ and $\mathrm{m}\textsubscript{2}$, the \textit{effective inspiral spin} ($\chi\textsubscript{eff}$) is given by
\begin{equation}
\chi\textsubscript{eff}=\frac{\mathrm{m}\textsubscript{1}\chi\textsubscript{1}+\mathrm{m}\textsubscript{2}\chi\textsubscript{2}}{\mathrm{m}\textsubscript{1}+\mathrm{m}\textsubscript{2}},
\end{equation}\label{eq:chieff}where $\chi\textsubscript{1}$ and $\chi\textsubscript{2}$ are the dimensionless component spins aligned to the orbital angular momentum of the binary. Its effect is imprinted in the signal emitted during the inspiral-merger-ringdown (IMR) stages of binary evolution. The \textit{signal-to-noise ratio} (SNR, or equivalently the loudness) of a signal in a detector is computed by maximizing the match between the recorded data ($d$) with a template waveform ($h$) modeling the signal. The SNR ($\rho$) in a detector is given by
\begin{equation}
\rho^{2}=\langle{d|h}\rangle=4\Re{\int_{f\textsubscript{lower}}^{f\textsubscript{upper}}\frac{\tilde{d}(f)\tilde{h}^{*}(f; \vect{\theta})}{\mathrm{S}\textsubscript{n}(f)}}df\,,
\end{equation}\label{eq:inner}where $\langle{d|h}\rangle$ represents the inner product (or the match) between the data and the template, with $\vect{\theta}$ describing the source of the signal. Here $\mathrm{S}\textsubscript{n}(f)$ represents the (one-sided) noise power spectral density (PSD) of the detector. The frequency interval bounded by $f\textsubscript{lower}$ and $f\textsubscript{upper}$ comes from the sensitive frequency band of the detector. For a detector network, the matched-filter network SNR is calculated as the quadrature sum of the SNRs in the detectors participating in the observation of an event. An empirical correlation between the inferred effective inspiral spin and the network SNR is shown in Fig.~\ref{fig:plotcomb}.
\begin{figure*}[t]
\vspace{-0.2cm}
\begin{centering}
\includegraphics[scale=0.78]{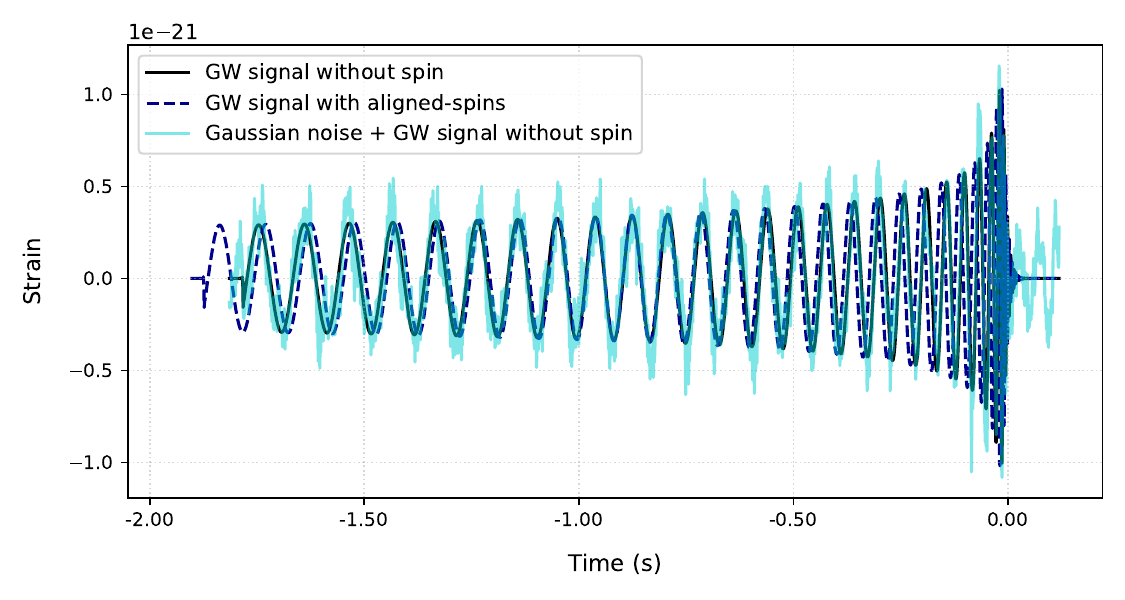}
\par\end{centering}
\vspace{-0.3cm}
\caption{Visualization of the effects of aligned-spins and that of the Gaussian noise on the GW strain in the time domain from an arbitrary simulated $(77.2, 57.6)$~$\mathrm{M}_{\odot}$ source located at $\sim$~1000~Mpc. The aligned-spins are such that $\chi\textsubscript{eff}=0.14$ (shown in dark blue). The GW strain for the non-spinning signal with a single realization of Gaussian noise is shown (in light blue).}\label{fig:visual}
\end{figure*}

In this work, inferences are mainly made from simulated astrophysical signals injected into Gaussian noise described as follows. The signal parameters are obtained from the injection sets presented with the public data release of \text{GWTC-4.0}~\citep{o4asens}. A subset of the injections within an arbitrary time-interval is selected. Since the properties of the underlying population do not vary with time, a sufficiently large time-interval to select injections is expected to preserve the population properties to a reasonable extent. A non-spinning injected population (or a corresponding population with aligned-spins only) is obtained by appropriately fixing the values of the spin parameters to zero for each injection in our injection set. Unless specified otherwise, full IMR signals are implemented using the \texttt{IMRPhenomXAS} waveform model~\citep{PhysRevD.102.064001}. The waveform can model GW signals from aligned-spin sources with an accuracy relevant for the current purpose. To simulate the data $d$, a signal ($s$) is deposited in Gaussian noise ($n$) obtained with detector PSDs available with the data release, which can be summarized as
\begin{equation}
d(t)=n(t)+s(t)\,.
\end{equation}\label{eq:data}This is illustrated in Fig.~\ref{fig:visual}. For demonstration, we simulate data for LIGO Hanford~(H1) and LIGO Livingston~(L1) and for ease of analysis, the injections are grouped into subsets categorized by the signal duration. Here we set the low-frequency cutoff for the analyses at $20$~Hz and the high-frequency cutoff is determined by at least the highest possible signal frequencies. For the simulations, $\chi\textsubscript{eff}$ values correspond to that obtained with a reference frequency set at the low-frequency cutoff.
\vspace{-0.2cm}
\section{Parameter inference}\label{sec:parest}
\begin{figure}[b]
\begin{centering}
\includegraphics[scale=0.65]{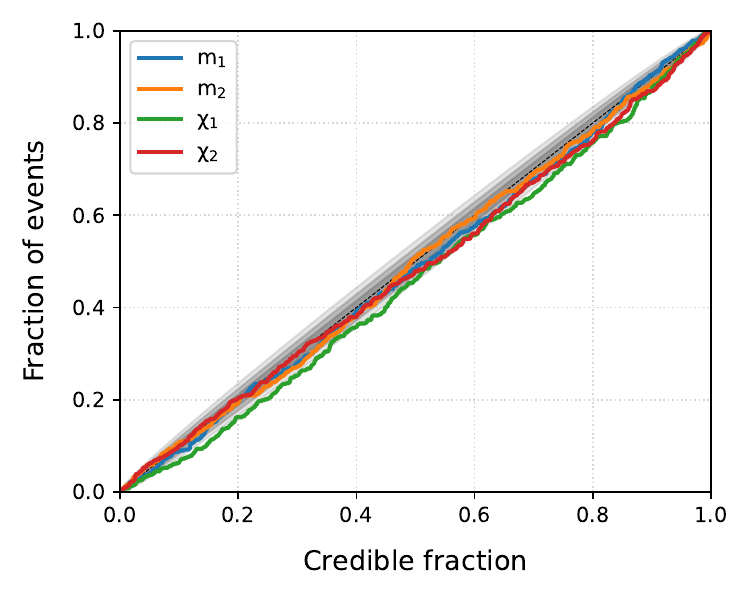}
\par\end{centering}
\vspace{-0.2cm}
\caption{Self-consistency of quoted credible intervals for the (intrinsic) source parameters used to calculate $\chi\textsubscript{eff}$. The grey shaded regions indicate up to $3\sigma$-uncertainty.}\label{fig:ppsam}
\end{figure}
\begin{figure*}[t]
\begin{centering}
\includegraphics[scale=0.48]{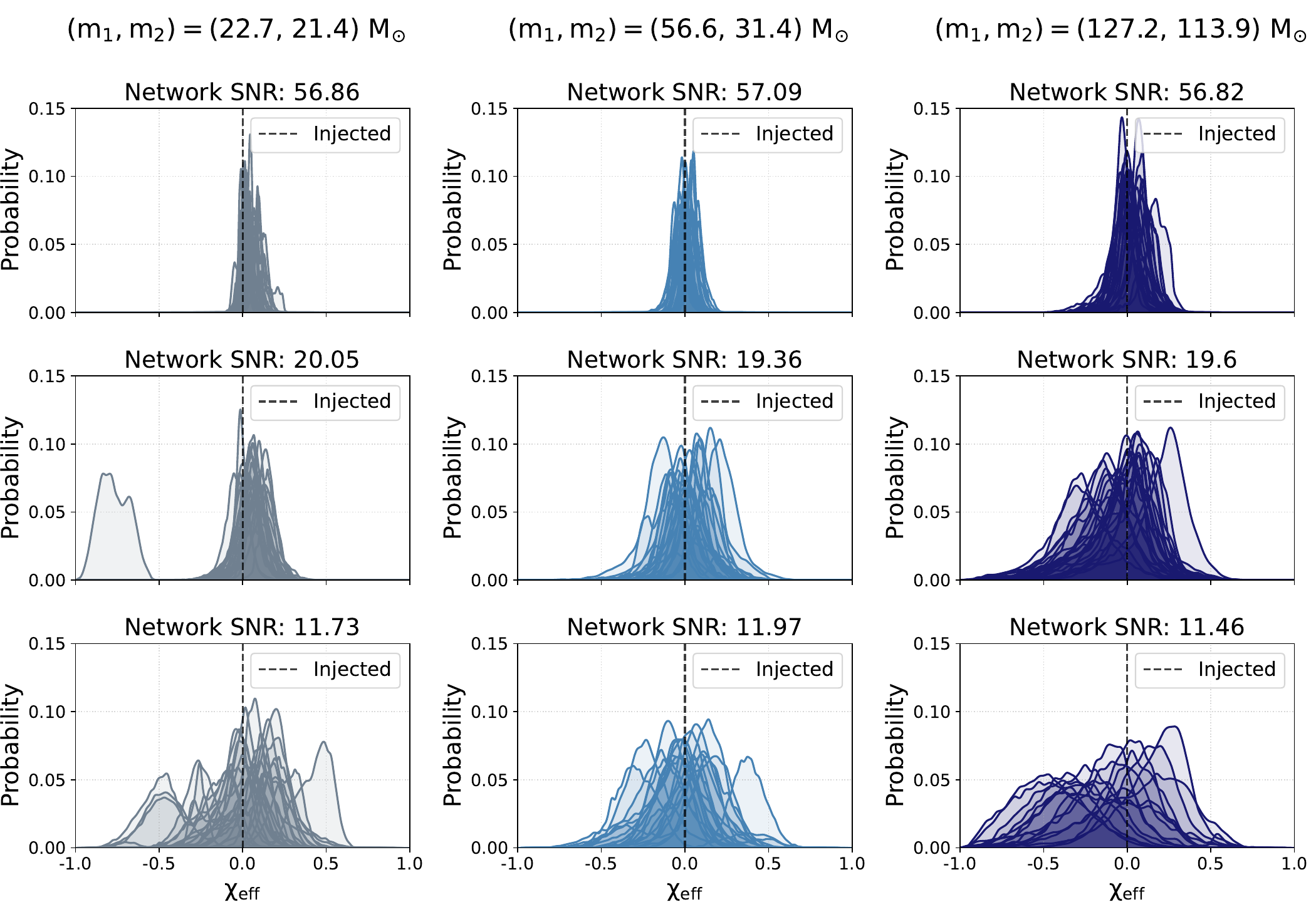}
\par\end{centering}
\vspace{-0.1cm}
\caption{Simulated effect of detector noise realizations on the inference of effective inspiral spin ($\chi\textsubscript{eff}$). Each column represents a given injected source with varied network SNRs. Each curve represents a unique noise realization for a given injected source. The network SNRs indicate the injected SNRs averaged over the multiple noise realizations.}\label{fig:plotnoirel}
\end{figure*}
To estimate the source parameters of an observed signal, we utilize Bayesian inference. Several algorithms are extensively described in the literature, see for example~\citep{lange2018rapid,Biwer_2019,Ashton_2019,PhysRevLett.127.241103,PhysRevD.108.082006,PhysRevD.108.064055,rml9-qyw1}. The goal is to assign a probability ($p$) to a range of possible values of any source parameter ($\theta$) given the data. The range of values is assumed to contain the true value. The process involves computation of the likelihood of observing the given data in the presence of a signal with trial/proposed parameter values that can possibly describe the source. For the underlying Gaussian noise, the likelihood of observing the data containing a signal with the parameters~$\vect{\theta}$ can be expressed as
\begin{equation}
{p}(d\,|\,\vect{\theta})\sim{e^{-\langle{d-h(\vect{\theta})|d-h(\vect{\theta})}\rangle/2}}.
\end{equation}\label{eq:likelihood}A prior assumption of the distribution of the probabilities, $p(\theta)$ is allowed to inform the reconstruction of the parameter~$\theta$. The reconstructed posterior probability can be represented as
\begin{equation}
p(\theta\,|\,d)\sim{p}(d\,|\,\theta)\:p(\theta).
\end{equation}\label{eq:probability}Here we choose to use astrophysically agnostic priors. In particular, for the component masses and spins, uniform priors in the respective ranges are used. The ranges are wide enough to contain the signal parameters of any given injection. The proposals for sampling the likelihoods, and hence the posterior probability distribution, are iteratively obtained using a stochastic algorithm. We use the \texttt{dynesty} sampler~\citep{joshspeagle} configured with \texttt{nlive}~$\sim$~1000 and \texttt{dlogZ}~$\sim$~0.1. The probability distribution of the parameter of interest can be obtained by marginalizing the posterior probabilities of all but the given parameter.

Here for any given event, the value of the parameter with the maximum probability represents its \textit{estimated} value. To obtain the uncertainty of the estimation, a highest probability neighbourhood of the estimated value is evaluated. For example,  for a 90\% \textit{credible} interval, the area of the distribution is summed up to 0.9 and the resulting interval is quoted\footnote{While a small uncertainty is desirable at any credibility, the sampling algorithm must not systematically deflate or inflate the probability distributions.}. The self-consistency of such intervals at any given credibility is validated in a probability-probability (pp-) plot. This is shown in Fig.~\ref{fig:ppsam} for the relevant source parameters using a set of aligned-spin injections. The diagonal nature of the plot indicates that the frequency of the credible intervals containing the true value of the parameter on the set of simulated sources is self-consistent on average. For sanity of the results, while the validation is necessary, note that it is obtained using single realization of noise segments in which the injected signals are deposited. In the next section, we discuss the possible effect of the noise realization on the measurement for any particular event.
\vspace{-0.2cm}
\section{Noise realization}\label{sec:noirel}
\begin{figure*}[t]
\vspace{-0.2cm}
\begin{centering}
\includegraphics[scale=0.69]{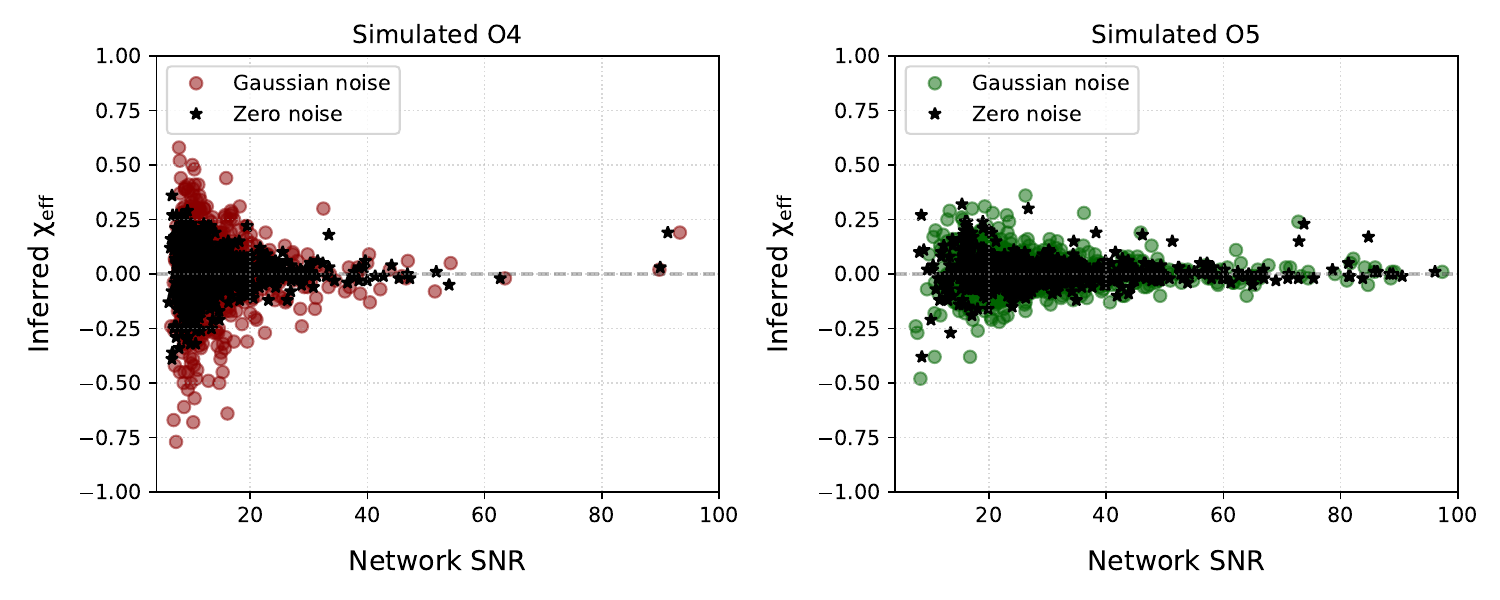}
\par\end{centering}
\vspace{-0.2cm}
\caption{Distribution of the estimated $\chi\textsubscript{eff}$ with the network SNR for a subset of non-spinning injections in Gaussian noise and in zero noise. The zero noise case indicates the ideal (or a measure of the best-achievable) estimate. A distinct noise seed is used for each injection in the Gaussian noise case. We observe that fainter events can be inferred with relatively larger non-zero $\chi\textsubscript{eff}$'s. We also note a larger scatter in the estimated values in the presence of noise. The analysis is repeated for the same set of injections, and hence for the same population of the observable sources, in the upcoming observing scenarios. It suggests that the effect is expected to be reduced in the O5-like observing scenario and beyond, given the overall greater SNRs recovered for the detected sources. The credible intervals are not shown here for visual clarity but they generally scale with the loudness.}\label{fig:gausvszero}
\end{figure*}

When an astrophysical signal arrives at a detector, the signal gets superimposed with the noise present in the detector. This is intrinsically a random, irreversible effect in the sense that the astrophysical signal cannot be separated from the noise or be reproduced with another realization. Thus, the likelihood of observing the data having a signal from a given source has an inherent terrestrial component coupled through the noise. So the inference on the observed astrophysical signal in the detector can possibly be affected by the noise component itself. The extent of this component on a given inference is expected to primarily depend on the loudness of the astrophysical event and the information content of the physical effect being probed. In other words, the impact is expected to be small if the physical effect induces a contribution to the likelihood of the data significantly more than the fluctuation caused by a specific noise realization. It can further depend on the randomness of a given realization, whether summarily in-phase, out-of-phase, or neutral to the physical effect.
\begin{figure}[b]
\begin{centering}
\includegraphics[scale=0.69]{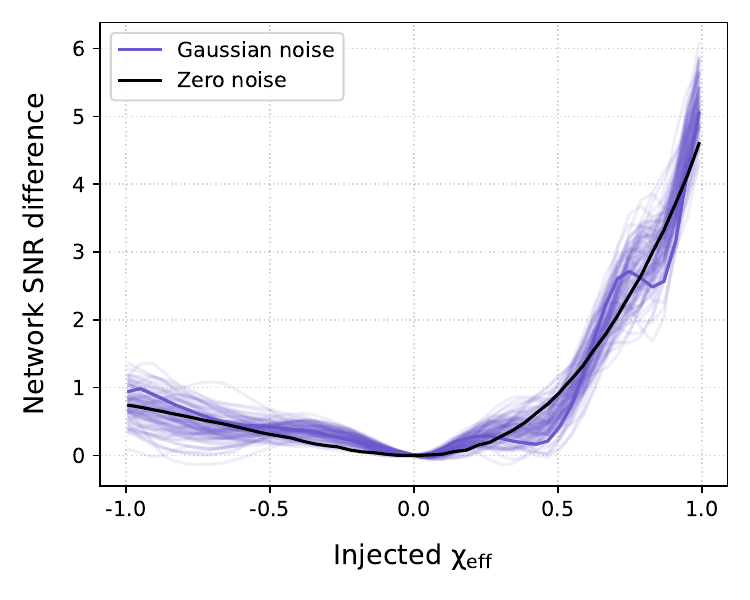}
\par\end{centering}
\vspace{-0.2cm}
\caption{Difference between aligned-spin SNRs and non-spinning SNRs plotted for different injected $\chi\textsubscript{eff}$ for a given injection with component masses ($85.6, 77.4$)~$\mathrm{M_{\odot}}$. The differences are compared for the injected signal in zero noise (in black) and in several Gaussian noise realizations (in blue); an arbitrary one is highlighted. We note that, on either side of the injected zero $\chi\textsubscript{eff}$, while the difference is strictly monotonic for zero noise, the same is ambiguous in the presence of noise, i.e. generally unpredictable depending on a specific realization of noise.}\label{fig:zeronoise}
\end{figure}
\begin{figure*}[t]
\vspace{-0.2cm}
\begin{centering}
\includegraphics[scale=0.75]{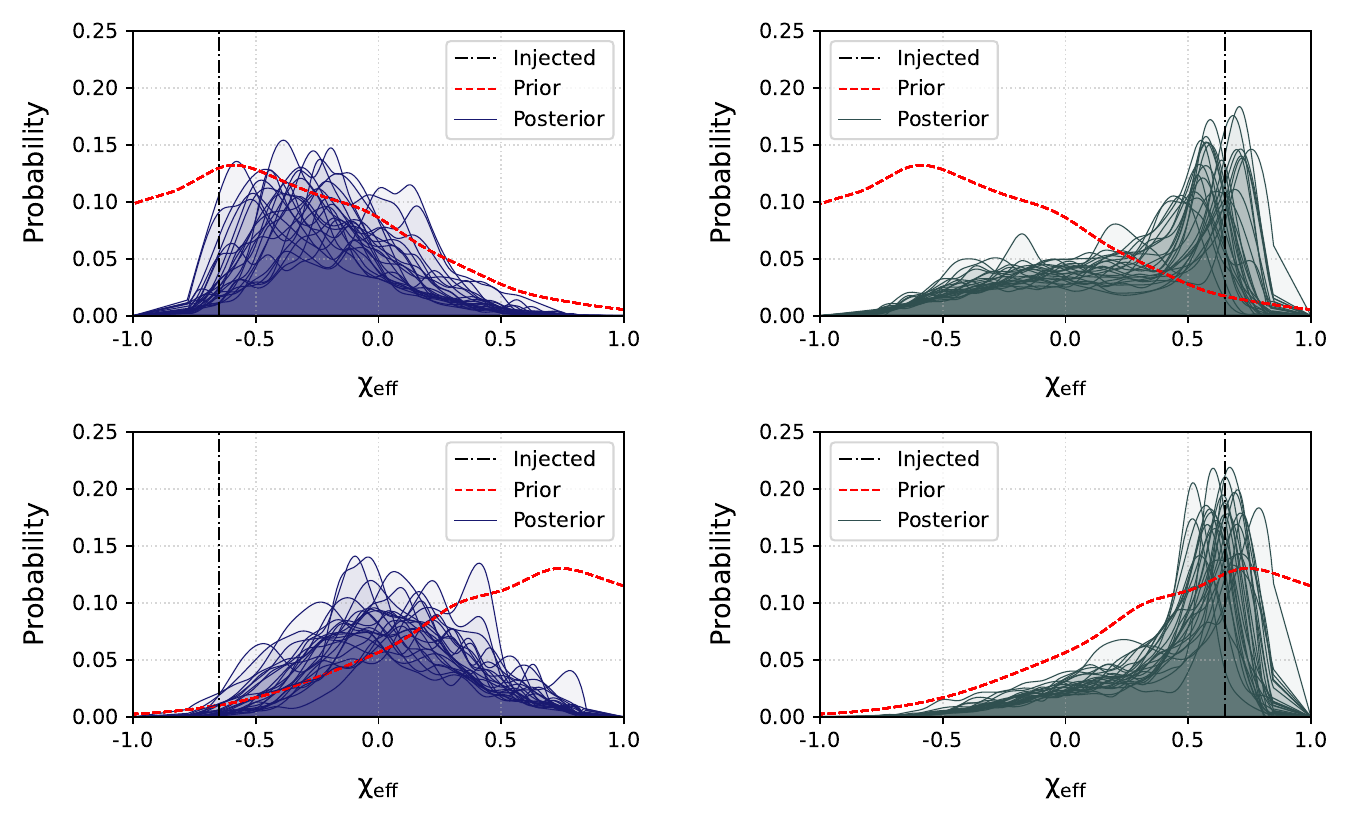}
\par\end{centering}
\vspace{-0.3cm}
\caption{Illustration of the effect of prior on the inference of $\chi\textsubscript{eff}$. For an injection with an arbitrary negative (left) and positive (right) $\chi\textsubscript{eff}$, the effect of two different priors in the presence of noise are shown. The prior distributions in $\chi\textsubscript{eff}$ result from Gaussian distributions in the aligned-spin components having a negative (top) and a positive (bottom) mean. The posterior distributions are not expected to be readily self-consistent with the use of such specialized priors. Note that the priors need not be astrophysically motivated for the illustrative purpose here.}\label{fig:prioreff}
\end{figure*}

Fortunately, the possible effect of noise realizations on an inference can be studied easily through simulations. Noise segments generated on a computer can be reproduced or newly generated altogether by specifying the random seed. Noise properties and attributes, such as the PSD, the segment duration, and the time-resolution, etc., are kept fixed. Thus, we conduct the inference of the source parameters of a simulated signal injected into multiple realizations of Gaussian noise. The injected luminosity distance to the source is varied to observe the effect at different network SNRs. They are shown in Fig.~\ref{fig:plotnoirel} for a few randomly selected non-spinning injections. For each of these cases, the posterior probability distributions for around 25 different noise realizations are shown. We note the shift in the peak of the distributions (i.e. the estimated values) from the injected value depending upon the SNRs of the signals. The resulting trend for a set of non-spinning injections is further shown in Fig.~\ref{fig:gausvszero} for O4 and O5-like observing scenarios. We also observe a similar trend in the inferred $\chi\textsubscript{eff}$ values of GW events from \text{GWTC-4.0+}, as shown earlier in Fig.~\ref{fig:plotcomb}. This effect is separate from and in addition to the broadening of the posterior distribution, thereby increasing the credible intervals at small signal strengths in general, which is indicated in Fig.~\ref{fig:plotnoirel}. A standard pp-plot obtained for a population of simulated sources is not expected to capture such effects since it is validated for a single realization of noise for each simulated signal. We further observe that at an approximate given SNR, heavier systems with smaller duration in the sensitive frequency band are typically more prone to this effect. For computational efficiency, we primarily conduct the analysis on simulated binary BH systems but the simulation framework is applicable to generic compact binary mergers, including spinning sources.

Parameter inferences often tend to depend on the choice of prior distribution used in the analyses, for example in determining the effective precession spin~\citep{ligo2025gwtc}. Ideally, with a sufficient sampling of the parameter space, a wide-enough prior distribution is expected to consistently reconstruct the posterior distribution\footnote{An inference may not be deemed robust if it is still largely dependent on the analysis configuration, including the priors.}. We argue that such dependencies can possibly occur since numerical algorithms can confuse themselves over any ambiguity in finding the maximum likelihood, or equivalently the network SNR. To test the possibility, we compute the difference in the estimated SNRs using an aligned-spin analysis and an otherwise same non-spinning analysis of a simulated signal, as shown in Fig.~\ref{fig:zeronoise}. We observe that while for the zero noise case, the SNR difference is approximately monotonic, the same is ambiguous for Gaussian noise cases. The dependence of an inference on the ambiguity is expected to be small when the physical effect being probed causes a large difference in the likelihood, beyond the fluctuations from the noise origin. We further note a dissymmetry in the SNR difference about the zero injected $\chi\textsubscript{eff}$. This generally indicates that the signal from a source with large aligned-spins carry more information as opposed to one with the corresponding magnitude of anti-aligned spins. We briefly discuss a potential implication below but do not explore its origin further here.

Here we expect that a variation in the initial (prior) distribution of samples can possibly favor a certain inference over another. To test this, we conduct the aligned-spin analysis of the given injection using two different prior choices. We use two different injected $\chi\textsubscript{eff}$ values (differing in signs but having the same magnitude) in the presence of Gaussian noise to demonstrate the effect described earlier. This is illustrated in Fig.~\ref{fig:prioreff}. We observe the average shift in the estimated $\chi\textsubscript{eff}$ and the associated credible intervals for the negative injected $\chi\textsubscript{eff}$ case with the different priors shown, though the priors are wide enough to accomodate the injected $\chi\textsubscript{eff}$ value. We expect this effect to be more pronounced with the measurement of the effective precession spin, provided it has a milder effect on the likelihood of the data from the signal.
\section{Population distinguishability}\label{sec:popinf}
In the previous section, we have shown that inference on individual sources can lead to non-zero effective inspiral spins even if they are truly non-spinning. However, inference on a fairly large number of sources can possibly inform us about the nature of the population. In principle, for a thorough understanding of a population, the distribution of each source parameter over every other parameter needs to be accounted for. However, since we mainly focus on the measurability of $\chi\textsubscript{eff}$ only, we explore whether a spinning population and an equivalent non-spinning population can be readily distinguished based on their spin measurements and thus we skip a full-scale, multi-dimensional population analysis.

Note that to attempt a qualitative inference at the population scale, we use the injections from the representative set of the detectable sources described earlier. For simplicity, we perform the parameter inference described earlier on a subset of the injections recovered from Gaussian noise only. Hence we skip the process of mitigating any effects from transient disturbances of terrestrial origin (or glitches) which can overlap with the astrophysical signals in real data. Also, since the inference is on events mapped with simulated signals as described below, the chance contamination due to noise events is negligible.

Here the injections are configured in two categories, without spins and with aligned-spins, by selectively restricting the spin components. For each of these configurations, events are recovered using a simulated matched-filter search where the template parameters are identical to the injection parameters. An injection is considered to be recovered when there is a coincident event (with individual detector SNRs $\geq$ 5) and the event lies (within a small time-window~$\sim$~$1$ s) around the coalescence time of the injection.

\begin{figure}[t]
\begin{centering}
\includegraphics[scale=0.69]{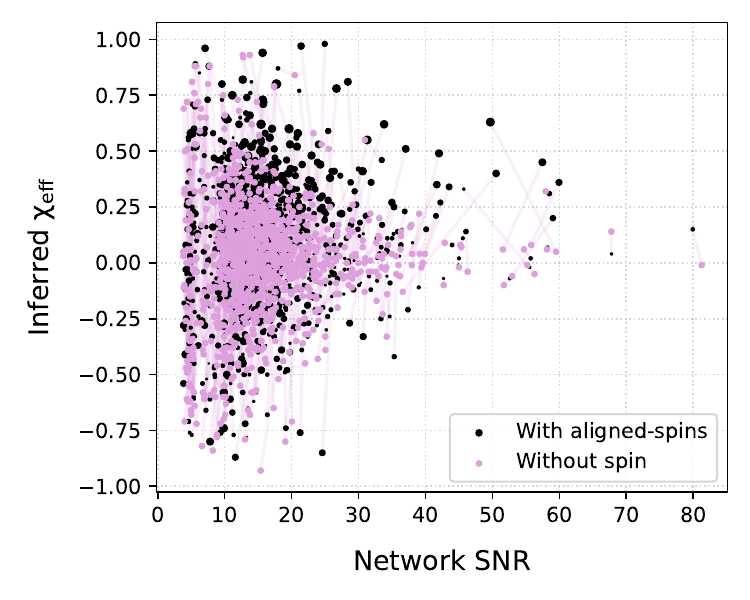}
\par\end{centering}
\vspace{-0.1cm}
\caption{Distinguishability of sources using their $\chi\textsubscript{eff}$ measurements on a set of aligned-spin injections and its equivalent set of non-spinning injections (mapped wherever available) in O4-like detector sensitivities. The size of data points (in black) represents the magnitude of the injected $\chi\textsubscript{eff}$. We observe no significant qualitative difference in the distributions of the inferred $\chi\textsubscript{eff}$ (typically at small magnitudes) for the two simulated populations at small to moderate SNRs.}\label{fig:astropop}
\end{figure}

We carry out parameter inference on the recovered injections for each configuration. The distribution of the inferred $\chi\textsubscript{eff}$ with the estimated network SNRs is plotted in Fig.~\ref{fig:astropop}. We do not observe any apparent primary distinction between the distributions for non-spinning and aligned-spinning populations, especially at small SNRs or at small magnitudes of inferred $\chi\textsubscript{eff}$. However, we note an excess in the number of events that are inferred with large $\chi\textsubscript{eff}$ and relatively large SNRs. Notably, for such events, the inferred $\chi\textsubscript{eff}$ for the corresponding non-spinning injections are typically close-to-zero. The abundance of the events, however, depends on the injected aligned-spin population. Such events are generally less abundant in the negative inferred $\chi\textsubscript{eff}$, which we attribute, in part, to the underlying behaviour of the inference described in Fig.~\ref{fig:zeronoise} and Fig.~\ref{fig:prioreff}. 

Given the effects from Gaussian noise alone, the inference on individual sources at small SNRs can systematically generate similar population-level effects for the two populations used here. Generally from this study, we cannot conclude with certainty whether the plethora of the reported GW events prefers any of the above populations, though by inspection, the distribution tends to be consistent with a population having minimal spins. We set expectations from a detailed analysis and a greater number of louder events in the detected population in the future.
\vspace{-0.2cm}
\section{Discussion}\label{sec:concl}
Through simulations, we have shown that a non-negligible effective inspiral spin can be inferred from a non-spinning compact binary merger with gravitational-wave observations. We found that the effect is generally pronounced for faint signals and that a random realization of Gaussian noise can enhance the effect. Some investigations are carried out to understand the underlying cause(s). We explored a probable cause to why an inference can be dependent on the choice of prior in a Bayesian analysis framework.

The possible effects of non-Gaussianity in the detector data on the inferences are not discussed here. To some extent glitches can be identified, modeled, and subtracted from the data before inferences are made, though several technical challenges exist arising primarily from the variety of the observed glitch types, see for example~\citep{PhysRevD.110.122002,PhysRevD.111.024046}. Thus, such effects are expected to be small after proper subtraction or mitigation and are secondary to the effects discussed in this work.

Importantly, the investigations point to the possibility of distorted outcomes in the measurement of any other astrophysical effects that are expected to induce small additional likelihood of the data in their presence, such as in the measurements of effective precession spin, the tidal deformability parameters in neutron star mergers, parameters that test gravity, etc. Appropriate investigations may be desirable while drawing conclusions based on parameter inference utilizing detector data. Generally, astrophysical inferences that are based on measurements on a small set of events could be statistically more prone to the effect.

This work also explores whether current estimates of $\chi\textsubscript{eff}$ can be readily used to distinguish an aligned-spin population from an equivalent non-spinning population. The population properties of the gravitational-wave sources observed by LIGO-Virgo are regularly explored and reported on a cumulative basis~\citep{Abbott_2021,PhysRevX.13.011048,ligo2025gwtcpop}. There are several efforts to distinguish any sub-populations from the inferred properties and/or accommodate all observed sources at the population-scale, see for example~\citep{szemraj2025disentangling,banagiri2025evidence,Galaudage_2021,adamcewicz2025both,PhysRevD.106.103019,Kimball_2020,Miller_2020,Callister_2022,PhysRevX.14.021005,PhysRevD.104.083010,PhysRevD.106.103019,Banagiri_2025,sadiq2025seeking,drsl-n3wz}. Here we report based on preliminary investigations that the current measurements can be consistent with a population of binary systems with minimal effective spins. The work also suggests that a greater number of events with large SNRs may be necessary to understand the true spin distribution in the observed population.

If compact binaries indeed have negligibly zero effective inspiral spin, then either each of the component aligned-spins must be essentially zero or the two mass-weighted aligned-spin components must cancel each other. Refer to a work that discusses intrinsic degeneracy effects~\citep{PhysRevD.93.084042}. We argue that when the estimation of effective spins are prone to the effects discussed in this work, it is further difficult to constrain individual component spins. It may be possible to confirm or rule out any such possibilities with adequate number of sufficiently loud observations. The work thus motivates for continued improvements in detector technologies and the pursuit for longer observation times, as have been happening in the past several decades. Future observations may be expected to be astrophysically illuminating of the population of the detected sources, see for example~\citep{Gupta_2024}, which can in turn help us better understand their formation process(es).
\vspace{0.35cm}
\section{Acknowledgments}\label{sec:ackno}
This research has made use of data or software obtained from the Gravitational Wave Open Science Center (gwosc.org), a service of the LIGO Scientific Collaboration, the Virgo Collaboration, and KAGRA. This material is based upon work supported by NSF's LIGO Laboratory which is a major facility fully funded by the National Science Foundation, as well as the Science and Technology Facilities Council (STFC) of the United Kingdom, the Max-Planck-Society (MPS), and the State of Niedersachsen/Germany for support of the construction of Advanced LIGO and construction and operation of the GEO600 detector. Additional support for Advanced LIGO was provided by the Australian Research Council. Virgo is funded, through the European Gravitational Observatory (EGO), by the French Centre National de Recherche Scientifique (CNRS), the Italian Istituto Nazionale di Fisica Nucleare (INFN) and the Dutch Nikhef, with contributions by institutions from Belgium, Germany, Greece, Hungary, Ireland, Japan, Monaco, Poland, Portugal, Spain. KAGRA is supported by Ministry of Education, Culture, Sports, Science and Technology (MEXT), Japan Society for the Promotion of Science (JSPS) in Japan; National Research Foundation (NRF) and Ministry of Science and ICT (MSIT) in Korea; Academia Sinica (AS) and National Science and Technology Council (NSTC) in Taiwan.

We have made use of the \texttt{PyCBC} software package for conducting various analyses~\citep{pycbcsoft}. The waveform models used in this work are publicly available as implemented in \texttt{LALSuite}~\citep{lalsuite}. We acknowledge the use of IUCAA-LDG cluster Sarathi for the computational/numerical work. The research was carried out at the Center of Excellence in Space Sciences India (CESSI). CESSI is a multi-institutional Center of Excellence hosted by the Indian Institute of Science Education and Research (IISER) Kolkata and has been established through funding from the Ministry of Education, Government of India. Due thanks to K.~Rajesh~Nayak. The author wishes to acknowledge feedback from the GW community. Sincere thanks to Lalit Pathak for carefully reading the manuscript. This document has the LIGO Document number P2500631. The author was financially self-supported.
\section{Data availability}\label{sec:sec1}
The data that support the findings of this article are openly available~\citep{o4apopzenodo,ligo2025open,gwosc}.
\bibliography{spin}

\begin{thebibliography}{43}%
\makeatletter
\providecommand \@ifxundefined [1]{%
 \@ifx{#1\undefined}
}%
\providecommand \@ifnum [1]{%
 \ifnum #1\expandafter \@firstoftwo
 \else \expandafter \@secondoftwo
 \fi
}%
\providecommand \@ifx [1]{%
 \ifx #1\expandafter \@firstoftwo
 \else \expandafter \@secondoftwo
 \fi
}%
\providecommand \natexlab [1]{#1}%
\providecommand \enquote  [1]{``#1''}%
\providecommand \bibnamefont  [1]{#1}%
\providecommand \bibfnamefont [1]{#1}%
\providecommand \citenamefont [1]{#1}%
\providecommand \href@noop [0]{\@secondoftwo}%
\providecommand \href [0]{\begingroup \@sanitize@url \@href}%
\providecommand \@href[1]{\@@startlink{#1}\@@href}%
\providecommand \@@href[1]{\endgroup#1\@@endlink}%
\providecommand \@sanitize@url [0]{\catcode `\\12\catcode `\$12\catcode
  `\&12\catcode `\#12\catcode `\^12\catcode `\_12\catcode `\%12\relax}%
\providecommand \@@startlink[1]{}%
\providecommand \@@endlink[0]{}%
\providecommand \url  [0]{\begingroup\@sanitize@url \@url }%
\providecommand \@url [1]{\endgroup\@href {#1}{\urlprefix }}%
\providecommand \urlprefix  [0]{URL }%
\providecommand \Eprint [0]{\href }%
\providecommand \doibase [0]{http://dx.doi.org/}%
\providecommand \selectlanguage [0]{\@gobble}%
\providecommand \bibinfo  [0]{\@secondoftwo}%
\providecommand \bibfield  [0]{\@secondoftwo}%
\providecommand \translation [1]{[#1]}%
\providecommand \BibitemOpen [0]{}%
\providecommand \bibitemStop [0]{}%
\providecommand \bibitemNoStop [0]{.\EOS\space}%
\providecommand \EOS [0]{\spacefactor3000\relax}%
\providecommand \BibitemShut  [1]{\csname bibitem#1\endcsname}%
\let\auto@bib@innerbib\@empty
%</preamble>
\bibitem [{\citenamefont {Abbott}\ \emph {et~al.}(2016)\citenamefont {Abbott},
  \citenamefont {Abbott}, \citenamefont {Abbott}, \citenamefont {Abernathy},
  \citenamefont {Acernese}, \citenamefont {Ackley}, \citenamefont {Adams},
  \citenamefont {Adams}, \citenamefont {Addesso}, \citenamefont {Adhikari}
  \emph {et~al.}}]{abbott2016observation}%
  \BibitemOpen
  \bibfield  {author} {\bibinfo {author} {\bibfnamefont {B.~P.}\ \bibnamefont
  {Abbott}}, \bibinfo {author} {\bibfnamefont {R.}~\bibnamefont {Abbott}},
  \bibinfo {author} {\bibfnamefont {T.}~\bibnamefont {Abbott}}, \bibinfo
  {author} {\bibfnamefont {M.}~\bibnamefont {Abernathy}}, \bibinfo {author}
  {\bibfnamefont {F.}~\bibnamefont {Acernese}}, \bibinfo {author}
  {\bibfnamefont {K.}~\bibnamefont {Ackley}}, \bibinfo {author} {\bibfnamefont
  {C.}~\bibnamefont {Adams}}, \bibinfo {author} {\bibfnamefont
  {T.}~\bibnamefont {Adams}}, \bibinfo {author} {\bibfnamefont
  {P.}~\bibnamefont {Addesso}}, \bibinfo {author} {\bibfnamefont
  {R.}~\bibnamefont {Adhikari}},  \emph {et~al.},\ }\href {\doibase
  10.1103/PhysRevLett.116.061102} {\bibfield  {journal} {\bibinfo  {journal}
  {Physical review letters}\ }\textbf {\bibinfo {volume} {116}},\ \bibinfo
  {pages} {061102} (\bibinfo {year} {2016})}\BibitemShut {NoStop}%
\bibitem [{\citenamefont {Collaboration}\ \emph
  {et~al.}(2025{\natexlab{a}})\citenamefont {Collaboration}, \citenamefont
  {Collaboration}, \citenamefont {Collaboration} \emph
  {et~al.}}]{ligo2025gwtc}%
  \BibitemOpen
  \bibfield  {author} {\bibinfo {author} {\bibfnamefont {L.~S.}\ \bibnamefont
  {Collaboration}}, \bibinfo {author} {\bibfnamefont {V.}~\bibnamefont
  {Collaboration}}, \bibinfo {author} {\bibfnamefont {K.}~\bibnamefont
  {Collaboration}},  \emph {et~al.},\ }\href {\doibase
  https://doi.org/10.48550/arXiv.2508.18082} {\bibfield  {journal} {\bibinfo
  {journal} {arXiv:2508.18082}\ } (\bibinfo {year} {2025}{\natexlab{a}}),\
  https://doi.org/10.48550/arXiv.2508.18082}\BibitemShut {NoStop}%
\bibitem [{\citenamefont {Abac}\ \emph
  {et~al.}(2025{\natexlab{a}})\citenamefont {Abac}, \citenamefont
  {Abouelfettouh}, \citenamefont {Acernese}, \citenamefont {Ackley} \emph
  {et~al.}}]{kw5gd732}%
  \BibitemOpen
  \bibfield  {author} {\bibinfo {author} {\bibfnamefont {A.~G.}\ \bibnamefont
  {Abac}}, \bibinfo {author} {\bibfnamefont {I.}~\bibnamefont {Abouelfettouh}},
  \bibinfo {author} {\bibfnamefont {F.}~\bibnamefont {Acernese}}, \bibinfo
  {author} {\bibfnamefont {K.}~\bibnamefont {Ackley}},  \emph {et~al.}
  (\bibinfo {collaboration} {LIGO Scientific, Virgo, and KAGRA
  Collaborations}),\ }\href {\doibase 10.1103/kw5g-d732} {\bibfield  {journal}
  {\bibinfo  {journal} {Phys. Rev. Lett.}\ }\textbf {\bibinfo {volume} {135}},\
  \bibinfo {pages} {111403} (\bibinfo {year} {2025}{\natexlab{a}})}\BibitemShut
  {NoStop}%
\bibitem [{\citenamefont {Abac}\ \emph
  {et~al.}(2025{\natexlab{b}})\citenamefont {Abac}, \citenamefont
  {Abouelfettouh}, \citenamefont {Acernese} \emph {et~al.}}]{Abac_2025}%
  \BibitemOpen
  \bibfield  {author} {\bibinfo {author} {\bibfnamefont {A.~G.}\ \bibnamefont
  {Abac}}, \bibinfo {author} {\bibfnamefont {I.}~\bibnamefont {Abouelfettouh}},
  \bibinfo {author} {\bibfnamefont {F.}~\bibnamefont {Acernese}},  \emph
  {et~al.} (\bibinfo {collaboration} {LIGO Scientific, Virgo, and KAGRA
  Collaborations}),\ }\href {\doibase 10.3847/2041-8213/ae0d54} {\bibfield
  {journal} {\bibinfo  {journal} {The Astrophysical Journal Letters}\ }\textbf
  {\bibinfo {volume} {993}},\ \bibinfo {pages} {L21} (\bibinfo {year}
  {2025}{\natexlab{b}})}\BibitemShut {NoStop}%
\bibitem [{\citenamefont {Abbott}\ \emph {et~al.}(2017)\citenamefont {Abbott},
  \citenamefont {Abbott}, \citenamefont {Abbott}, \citenamefont {Acernese},
  \citenamefont {Ackley}, \citenamefont {Adams}, \citenamefont {Adams},
  \citenamefont {Addesso}, \citenamefont {Adhikari}, \citenamefont {Adya} \emph
  {et~al.}}]{abbott2017gw170817}%
  \BibitemOpen
  \bibfield  {author} {\bibinfo {author} {\bibfnamefont {B.~P.}\ \bibnamefont
  {Abbott}}, \bibinfo {author} {\bibfnamefont {R.}~\bibnamefont {Abbott}},
  \bibinfo {author} {\bibfnamefont {T.}~\bibnamefont {Abbott}}, \bibinfo
  {author} {\bibfnamefont {F.}~\bibnamefont {Acernese}}, \bibinfo {author}
  {\bibfnamefont {K.}~\bibnamefont {Ackley}}, \bibinfo {author} {\bibfnamefont
  {C.}~\bibnamefont {Adams}}, \bibinfo {author} {\bibfnamefont
  {T.}~\bibnamefont {Adams}}, \bibinfo {author} {\bibfnamefont
  {P.}~\bibnamefont {Addesso}}, \bibinfo {author} {\bibfnamefont
  {R.}~\bibnamefont {Adhikari}}, \bibinfo {author} {\bibfnamefont {V.~B.}\
  \bibnamefont {Adya}},  \emph {et~al.},\ }\href {\doibase
  10.1103/PhysRevLett.119.161101} {\bibfield  {journal} {\bibinfo  {journal}
  {Physical review letters}\ }\textbf {\bibinfo {volume} {119}},\ \bibinfo
  {pages} {161101} (\bibinfo {year} {2017})}\BibitemShut {NoStop}%
\bibitem [{\citenamefont {Abbott}\ \emph {et~al.}(2020)\citenamefont {Abbott},
  \citenamefont {Abbott}, \citenamefont {Abbott}, \citenamefont {Abraham},
  \citenamefont {Acernese},\ and\ \citenamefont {Ackley}}]{Abbott_2020}%
  \BibitemOpen
  \bibfield  {author} {\bibinfo {author} {\bibfnamefont {B.~P.}\ \bibnamefont
  {Abbott}}, \bibinfo {author} {\bibfnamefont {R.}~\bibnamefont {Abbott}},
  \bibinfo {author} {\bibfnamefont {T.~D.}\ \bibnamefont {Abbott}}, \bibinfo
  {author} {\bibfnamefont {S.}~\bibnamefont {Abraham}}, \bibinfo {author}
  {\bibfnamefont {F.}~\bibnamefont {Acernese}}, \ and\ \bibinfo {author}
  {\bibfnamefont {K.}~\bibnamefont {Ackley}},\ }\href {\doibase
  10.3847/2041-8213/ab75f5} {\bibfield  {journal} {\bibinfo  {journal} {The
  Astrophysical Journal Letters}\ }\textbf {\bibinfo {volume} {892}},\ \bibinfo
  {pages} {L3} (\bibinfo {year} {2020})}\BibitemShut {NoStop}%
\bibitem [{\citenamefont {Abbott}\ \emph
  {et~al.}(2021{\natexlab{a}})\citenamefont {Abbott}, \citenamefont {Abbott},
  \citenamefont {Abraham}, \citenamefont {Acernese}, \citenamefont {Ackley},
  \citenamefont {Adams}, \citenamefont {Adams}, \citenamefont {Adhikari},
  \citenamefont {Adya}, \citenamefont {Affeldt} \emph
  {et~al.}}]{abbott2021observation}%
  \BibitemOpen
  \bibfield  {author} {\bibinfo {author} {\bibfnamefont {R.}~\bibnamefont
  {Abbott}}, \bibinfo {author} {\bibfnamefont {T.}~\bibnamefont {Abbott}},
  \bibinfo {author} {\bibfnamefont {S.}~\bibnamefont {Abraham}}, \bibinfo
  {author} {\bibfnamefont {F.}~\bibnamefont {Acernese}}, \bibinfo {author}
  {\bibfnamefont {K.}~\bibnamefont {Ackley}}, \bibinfo {author} {\bibfnamefont
  {A.}~\bibnamefont {Adams}}, \bibinfo {author} {\bibfnamefont
  {C.}~\bibnamefont {Adams}}, \bibinfo {author} {\bibfnamefont
  {R.}~\bibnamefont {Adhikari}}, \bibinfo {author} {\bibfnamefont
  {V.}~\bibnamefont {Adya}}, \bibinfo {author} {\bibfnamefont {C.}~\bibnamefont
  {Affeldt}},  \emph {et~al.},\ }\href {\doibase 10.3847/2041-8213/ac082e}
  {\bibfield  {journal} {\bibinfo  {journal} {The Astrophysical journal
  letters}\ }\textbf {\bibinfo {volume} {915}},\ \bibinfo {pages} {L5}
  (\bibinfo {year} {2021}{\natexlab{a}})}\BibitemShut {NoStop}%
\bibitem [{gwo({\natexlab{a}})}]{gwoscevents}%
  \BibitemOpen
  \href@noop {} {}\bibinfo {howpublished}
  {\url{https://gwosc.org/eventapi/html/allevents/}}
  ({\natexlab{a}})\BibitemShut {NoStop}%
\bibitem [{o4a({\natexlab{a}})}]{o4asens}%
  \BibitemOpen
  \href@noop {} {}\bibinfo {howpublished}
  {\url{https://doi.org/10.5281/zenodo.16740117}} ({\natexlab{a}})\BibitemShut
  {NoStop}%
\bibitem [{\citenamefont {Pratten}\ \emph {et~al.}(2020)\citenamefont
  {Pratten}, \citenamefont {Husa}, \citenamefont {Garc\'{\i}a-Quir\'os},
  \citenamefont {Colleoni}, \citenamefont {Ramos-Buades}, \citenamefont
  {Estell\'es},\ and\ \citenamefont {Jaume}}]{PhysRevD.102.064001}%
  \BibitemOpen
  \bibfield  {author} {\bibinfo {author} {\bibfnamefont {G.}~\bibnamefont
  {Pratten}}, \bibinfo {author} {\bibfnamefont {S.}~\bibnamefont {Husa}},
  \bibinfo {author} {\bibfnamefont {C.}~\bibnamefont {Garc\'{\i}a-Quir\'os}},
  \bibinfo {author} {\bibfnamefont {M.}~\bibnamefont {Colleoni}}, \bibinfo
  {author} {\bibfnamefont {A.}~\bibnamefont {Ramos-Buades}}, \bibinfo {author}
  {\bibfnamefont {H.}~\bibnamefont {Estell\'es}}, \ and\ \bibinfo {author}
  {\bibfnamefont {R.}~\bibnamefont {Jaume}},\ }\href {\doibase
  10.1103/PhysRevD.102.064001} {\bibfield  {journal} {\bibinfo  {journal}
  {Phys. Rev. D}\ }\textbf {\bibinfo {volume} {102}},\ \bibinfo {pages}
  {064001} (\bibinfo {year} {2020})}\BibitemShut {NoStop}%
\bibitem [{\citenamefont {Lange}\ \emph {et~al.}(2018)\citenamefont {Lange},
  \citenamefont {O'Shaughnessy},\ and\ \citenamefont {Rizzo}}]{lange2018rapid}%
  \BibitemOpen
  \bibfield  {author} {\bibinfo {author} {\bibfnamefont {J.}~\bibnamefont
  {Lange}}, \bibinfo {author} {\bibfnamefont {R.}~\bibnamefont
  {O'Shaughnessy}}, \ and\ \bibinfo {author} {\bibfnamefont {M.}~\bibnamefont
  {Rizzo}},\ }\href {\doibase https://doi.org/10.48550/arXiv.1805.10457}
  {\bibfield  {journal} {\bibinfo  {journal} {arXiv:1805.10457}\ } (\bibinfo
  {year} {2018}),\ https://doi.org/10.48550/arXiv.1805.10457}\BibitemShut
  {NoStop}%
\bibitem [{\citenamefont {Biwer}\ \emph {et~al.}(2019)\citenamefont {Biwer},
  \citenamefont {Capano}, \citenamefont {De}, \citenamefont {Cabero},
  \citenamefont {Brown}, \citenamefont {Nitz},\ and\ \citenamefont
  {Raymond}}]{Biwer_2019}%
  \BibitemOpen
  \bibfield  {author} {\bibinfo {author} {\bibfnamefont {C.~M.}\ \bibnamefont
  {Biwer}}, \bibinfo {author} {\bibfnamefont {C.~D.}\ \bibnamefont {Capano}},
  \bibinfo {author} {\bibfnamefont {S.}~\bibnamefont {De}}, \bibinfo {author}
  {\bibfnamefont {M.}~\bibnamefont {Cabero}}, \bibinfo {author} {\bibfnamefont
  {D.~A.}\ \bibnamefont {Brown}}, \bibinfo {author} {\bibfnamefont {A.~H.}\
  \bibnamefont {Nitz}}, \ and\ \bibinfo {author} {\bibfnamefont
  {V.}~\bibnamefont {Raymond}},\ }\href {\doibase 10.1088/1538-3873/aaef0b}
  {\bibfield  {journal} {\bibinfo  {journal} {Publications of the Astronomical
  Society of the Pacific}\ }\textbf {\bibinfo {volume} {131}},\ \bibinfo
  {pages} {024503} (\bibinfo {year} {2019})}\BibitemShut {NoStop}%
\bibitem [{\citenamefont {Ashton}\ \emph {et~al.}(2019)\citenamefont {Ashton},
  \citenamefont {Hübner}, \citenamefont {Lasky}, \citenamefont {Talbot},
  \citenamefont {Ackley}, \citenamefont {Biscoveanu}, \citenamefont {Chu},
  \citenamefont {Divakarla}, \citenamefont {Easter}, \citenamefont {Goncharov},
  \citenamefont {Vivanco}, \citenamefont {Harms}, \citenamefont {Lower},
  \citenamefont {Meadors}, \citenamefont {Melchor}, \citenamefont {Payne},
  \citenamefont {Pitkin}, \citenamefont {Powell}, \citenamefont {Sarin},
  \citenamefont {Smith},\ and\ \citenamefont {Thrane}}]{Ashton_2019}%
  \BibitemOpen
  \bibfield  {author} {\bibinfo {author} {\bibfnamefont {G.}~\bibnamefont
  {Ashton}}, \bibinfo {author} {\bibfnamefont {M.}~\bibnamefont {Hübner}},
  \bibinfo {author} {\bibfnamefont {P.~D.}\ \bibnamefont {Lasky}}, \bibinfo
  {author} {\bibfnamefont {C.}~\bibnamefont {Talbot}}, \bibinfo {author}
  {\bibfnamefont {K.}~\bibnamefont {Ackley}}, \bibinfo {author} {\bibfnamefont
  {S.}~\bibnamefont {Biscoveanu}}, \bibinfo {author} {\bibfnamefont
  {Q.}~\bibnamefont {Chu}}, \bibinfo {author} {\bibfnamefont {A.}~\bibnamefont
  {Divakarla}}, \bibinfo {author} {\bibfnamefont {P.~J.}\ \bibnamefont
  {Easter}}, \bibinfo {author} {\bibfnamefont {B.}~\bibnamefont {Goncharov}},
  \bibinfo {author} {\bibfnamefont {F.~H.}\ \bibnamefont {Vivanco}}, \bibinfo
  {author} {\bibfnamefont {J.}~\bibnamefont {Harms}}, \bibinfo {author}
  {\bibfnamefont {M.~E.}\ \bibnamefont {Lower}}, \bibinfo {author}
  {\bibfnamefont {G.~D.}\ \bibnamefont {Meadors}}, \bibinfo {author}
  {\bibfnamefont {D.}~\bibnamefont {Melchor}}, \bibinfo {author} {\bibfnamefont
  {E.}~\bibnamefont {Payne}}, \bibinfo {author} {\bibfnamefont {M.~D.}\
  \bibnamefont {Pitkin}}, \bibinfo {author} {\bibfnamefont {J.}~\bibnamefont
  {Powell}}, \bibinfo {author} {\bibfnamefont {N.}~\bibnamefont {Sarin}},
  \bibinfo {author} {\bibfnamefont {R.~J.~E.}\ \bibnamefont {Smith}}, \ and\
  \bibinfo {author} {\bibfnamefont {E.}~\bibnamefont {Thrane}},\ }\href
  {\doibase 10.3847/1538-4365/ab06fc} {\bibfield  {journal} {\bibinfo
  {journal} {The Astrophysical Journal Supplement Series}\ }\textbf {\bibinfo
  {volume} {241}},\ \bibinfo {pages} {27} (\bibinfo {year} {2019})}\BibitemShut
  {NoStop}%
\bibitem [{\citenamefont {Dax}\ \emph {et~al.}(2021)\citenamefont {Dax},
  \citenamefont {Green}, \citenamefont {Gair}, \citenamefont {Macke},
  \citenamefont {Buonanno},\ and\ \citenamefont
  {Sch\"olkopf}}]{PhysRevLett.127.241103}%
  \BibitemOpen
  \bibfield  {author} {\bibinfo {author} {\bibfnamefont {M.}~\bibnamefont
  {Dax}}, \bibinfo {author} {\bibfnamefont {S.~R.}\ \bibnamefont {Green}},
  \bibinfo {author} {\bibfnamefont {J.}~\bibnamefont {Gair}}, \bibinfo {author}
  {\bibfnamefont {J.~H.}\ \bibnamefont {Macke}}, \bibinfo {author}
  {\bibfnamefont {A.}~\bibnamefont {Buonanno}}, \ and\ \bibinfo {author}
  {\bibfnamefont {B.}~\bibnamefont {Sch\"olkopf}},\ }\href {\doibase
  10.1103/PhysRevLett.127.241103} {\bibfield  {journal} {\bibinfo  {journal}
  {Phys. Rev. Lett.}\ }\textbf {\bibinfo {volume} {127}},\ \bibinfo {pages}
  {241103} (\bibinfo {year} {2021})}\BibitemShut {NoStop}%
\bibitem [{\citenamefont {Fairhurst}\ \emph {et~al.}(2023)\citenamefont
  {Fairhurst}, \citenamefont {Hoy}, \citenamefont {Green}, \citenamefont
  {Mills},\ and\ \citenamefont {Usman}}]{PhysRevD.108.082006}%
  \BibitemOpen
  \bibfield  {author} {\bibinfo {author} {\bibfnamefont {S.}~\bibnamefont
  {Fairhurst}}, \bibinfo {author} {\bibfnamefont {C.}~\bibnamefont {Hoy}},
  \bibinfo {author} {\bibfnamefont {R.}~\bibnamefont {Green}}, \bibinfo
  {author} {\bibfnamefont {C.}~\bibnamefont {Mills}}, \ and\ \bibinfo {author}
  {\bibfnamefont {S.~A.}\ \bibnamefont {Usman}},\ }\href {\doibase
  10.1103/PhysRevD.108.082006} {\bibfield  {journal} {\bibinfo  {journal}
  {Phys. Rev. D}\ }\textbf {\bibinfo {volume} {108}},\ \bibinfo {pages}
  {082006} (\bibinfo {year} {2023})}\BibitemShut {NoStop}%
\bibitem [{\citenamefont {Pathak}\ \emph {et~al.}(2023)\citenamefont {Pathak},
  \citenamefont {Reza},\ and\ \citenamefont {Sengupta}}]{PhysRevD.108.064055}%
  \BibitemOpen
  \bibfield  {author} {\bibinfo {author} {\bibfnamefont {L.}~\bibnamefont
  {Pathak}}, \bibinfo {author} {\bibfnamefont {A.}~\bibnamefont {Reza}}, \ and\
  \bibinfo {author} {\bibfnamefont {A.~S.}\ \bibnamefont {Sengupta}},\ }\href
  {\doibase 10.1103/PhysRevD.108.064055} {\bibfield  {journal} {\bibinfo
  {journal} {Phys. Rev. D}\ }\textbf {\bibinfo {volume} {108}},\ \bibinfo
  {pages} {064055} (\bibinfo {year} {2023})}\BibitemShut {NoStop}%
\bibitem [{\citenamefont {Nitz}(2025)}]{rml9-qyw1}%
  \BibitemOpen
  \bibfield  {author} {\bibinfo {author} {\bibfnamefont {A.~H.}\ \bibnamefont
  {Nitz}},\ }\href {\doibase 10.1103/rml9-qyw1} {\bibfield  {journal} {\bibinfo
   {journal} {Phys. Rev. D}\ }\textbf {\bibinfo {volume} {112}},\ \bibinfo
  {pages} {023032} (\bibinfo {year} {2025})}\BibitemShut {NoStop}%
\bibitem [{jos()}]{joshspeagle}%
  \BibitemOpen
  \href@noop {} {}\bibinfo {howpublished}
  {\url{https://doi.org/10.5281/zenodo.17268284}}\BibitemShut {NoStop}%
\bibitem [{\citenamefont {Ghonge}\ \emph {et~al.}(2024)\citenamefont {Ghonge},
  \citenamefont {Brandt}, \citenamefont {Sullivan}, \citenamefont {Millhouse},
  \citenamefont {Chatziioannou}, \citenamefont {Clark}, \citenamefont
  {Littenberg}, \citenamefont {Cornish}, \citenamefont {Hourihane},\ and\
  \citenamefont {Cadonati}}]{PhysRevD.110.122002}%
  \BibitemOpen
  \bibfield  {author} {\bibinfo {author} {\bibfnamefont {S.}~\bibnamefont
  {Ghonge}}, \bibinfo {author} {\bibfnamefont {J.}~\bibnamefont {Brandt}},
  \bibinfo {author} {\bibfnamefont {J.~M.}\ \bibnamefont {Sullivan}}, \bibinfo
  {author} {\bibfnamefont {M.}~\bibnamefont {Millhouse}}, \bibinfo {author}
  {\bibfnamefont {K.}~\bibnamefont {Chatziioannou}}, \bibinfo {author}
  {\bibfnamefont {J.~A.}\ \bibnamefont {Clark}}, \bibinfo {author}
  {\bibfnamefont {T.}~\bibnamefont {Littenberg}}, \bibinfo {author}
  {\bibfnamefont {N.}~\bibnamefont {Cornish}}, \bibinfo {author} {\bibfnamefont
  {S.}~\bibnamefont {Hourihane}}, \ and\ \bibinfo {author} {\bibfnamefont
  {L.}~\bibnamefont {Cadonati}},\ }\href {\doibase 10.1103/PhysRevD.110.122002}
  {\bibfield  {journal} {\bibinfo  {journal} {Phys. Rev. D}\ }\textbf {\bibinfo
  {volume} {110}},\ \bibinfo {pages} {122002} (\bibinfo {year}
  {2024})}\BibitemShut {NoStop}%
\bibitem [{\citenamefont {Udall}\ \emph {et~al.}(2025)\citenamefont {Udall},
  \citenamefont {Hourihane}, \citenamefont {Miller}, \citenamefont {Davis},
  \citenamefont {Chatziioannou}, \citenamefont {Isi},\ and\ \citenamefont
  {Deshong}}]{PhysRevD.111.024046}%
  \BibitemOpen
  \bibfield  {author} {\bibinfo {author} {\bibfnamefont {R.}~\bibnamefont
  {Udall}}, \bibinfo {author} {\bibfnamefont {S.}~\bibnamefont {Hourihane}},
  \bibinfo {author} {\bibfnamefont {S.}~\bibnamefont {Miller}}, \bibinfo
  {author} {\bibfnamefont {D.}~\bibnamefont {Davis}}, \bibinfo {author}
  {\bibfnamefont {K.}~\bibnamefont {Chatziioannou}}, \bibinfo {author}
  {\bibfnamefont {M.}~\bibnamefont {Isi}}, \ and\ \bibinfo {author}
  {\bibfnamefont {H.}~\bibnamefont {Deshong}},\ }\href {\doibase
  10.1103/PhysRevD.111.024046} {\bibfield  {journal} {\bibinfo  {journal}
  {Phys. Rev. D}\ }\textbf {\bibinfo {volume} {111}},\ \bibinfo {pages}
  {024046} (\bibinfo {year} {2025})}\BibitemShut {NoStop}%
\bibitem [{\citenamefont {Abbott}\ \emph
  {et~al.}(2021{\natexlab{b}})\citenamefont {Abbott}, \citenamefont {Abbott},
  \citenamefont {Abraham} \emph {et~al.}}]{Abbott_2021}%
  \BibitemOpen
  \bibfield  {author} {\bibinfo {author} {\bibfnamefont {R.}~\bibnamefont
  {Abbott}}, \bibinfo {author} {\bibfnamefont {T.~D.}\ \bibnamefont {Abbott}},
  \bibinfo {author} {\bibfnamefont {S.}~\bibnamefont {Abraham}},  \emph
  {et~al.},\ }\href {\doibase 10.3847/2041-8213/abe949} {\bibfield  {journal}
  {\bibinfo  {journal} {The Astrophysical Journal Letters}\ }\textbf {\bibinfo
  {volume} {913}},\ \bibinfo {pages} {L7} (\bibinfo {year}
  {2021}{\natexlab{b}})}\BibitemShut {NoStop}%
\bibitem [{\citenamefont {Abbott}\ \emph {et~al.}(2023)\citenamefont {Abbott},
  \citenamefont {Abbott}, \citenamefont {Acernese} \emph
  {et~al.}}]{PhysRevX.13.011048}%
  \BibitemOpen
  \bibfield  {author} {\bibinfo {author} {\bibfnamefont {R.}~\bibnamefont
  {Abbott}}, \bibinfo {author} {\bibfnamefont {T.~D.}\ \bibnamefont {Abbott}},
  \bibinfo {author} {\bibfnamefont {F.}~\bibnamefont {Acernese}},  \emph
  {et~al.} (\bibinfo {collaboration} {LIGO Scientific Collaboration, Virgo
  Collaboration, and KAGRA Collaboration}),\ }\href {\doibase
  10.1103/PhysRevX.13.011048} {\bibfield  {journal} {\bibinfo  {journal} {Phys.
  Rev. X}\ }\textbf {\bibinfo {volume} {13}},\ \bibinfo {pages} {011048}
  (\bibinfo {year} {2023})}\BibitemShut {NoStop}%
\bibitem [{\citenamefont {Collaboration}\ \emph
  {et~al.}(2025{\natexlab{b}})\citenamefont {Collaboration}, \citenamefont
  {Collaboration}, \citenamefont {Collaboration} \emph
  {et~al.}}]{ligo2025gwtcpop}%
  \BibitemOpen
  \bibfield  {author} {\bibinfo {author} {\bibfnamefont {L.~S.}\ \bibnamefont
  {Collaboration}}, \bibinfo {author} {\bibfnamefont {V.}~\bibnamefont
  {Collaboration}}, \bibinfo {author} {\bibfnamefont {K.}~\bibnamefont
  {Collaboration}},  \emph {et~al.},\ }\href {\doibase
  https://doi.org/10.48550/arXiv.2508.18083} {\bibfield  {journal} {\bibinfo
  {journal} {arXiv:2508.18083}\ } (\bibinfo {year} {2025}{\natexlab{b}}),\
  https://doi.org/10.48550/arXiv.2508.18083}\BibitemShut {NoStop}%
\bibitem [{\citenamefont {Szemraj}\ and\ \citenamefont
  {Biscoveanu}(2025)}]{szemraj2025disentangling}%
  \BibitemOpen
  \bibfield  {author} {\bibinfo {author} {\bibfnamefont {L.}~\bibnamefont
  {Szemraj}}\ and\ \bibinfo {author} {\bibfnamefont {S.}~\bibnamefont
  {Biscoveanu}},\ }\href {\doibase https://doi.org/10.48550/arXiv.2507.23663}
  {\bibfield  {journal} {\bibinfo  {journal} {arXiv:2507.23663}\ } (\bibinfo
  {year} {2025}),\ https://doi.org/10.48550/arXiv.2507.23663}\BibitemShut
  {NoStop}%
\bibitem [{\citenamefont {Banagiri}\ \emph
  {et~al.}(2025{\natexlab{a}})\citenamefont {Banagiri}, \citenamefont
  {Thrane},\ and\ \citenamefont {Lasky}}]{banagiri2025evidence}%
  \BibitemOpen
  \bibfield  {author} {\bibinfo {author} {\bibfnamefont {S.}~\bibnamefont
  {Banagiri}}, \bibinfo {author} {\bibfnamefont {E.}~\bibnamefont {Thrane}}, \
  and\ \bibinfo {author} {\bibfnamefont {P.~D.}\ \bibnamefont {Lasky}},\ }\href
  {\doibase https://doi.org/10.48550/arXiv.2509.15646} {\bibfield  {journal}
  {\bibinfo  {journal} {arXiv:2509.15646}\ } (\bibinfo {year}
  {2025}{\natexlab{a}}),\
  https://doi.org/10.48550/arXiv.2509.15646}\BibitemShut {NoStop}%
\bibitem [{\citenamefont {Galaudage}\ \emph {et~al.}(2021)\citenamefont
  {Galaudage}, \citenamefont {Talbot}, \citenamefont {Nagar}, \citenamefont
  {Jain}, \citenamefont {Thrane},\ and\ \citenamefont
  {Mandel}}]{Galaudage_2021}%
  \BibitemOpen
  \bibfield  {author} {\bibinfo {author} {\bibfnamefont {S.}~\bibnamefont
  {Galaudage}}, \bibinfo {author} {\bibfnamefont {C.}~\bibnamefont {Talbot}},
  \bibinfo {author} {\bibfnamefont {T.}~\bibnamefont {Nagar}}, \bibinfo
  {author} {\bibfnamefont {D.}~\bibnamefont {Jain}}, \bibinfo {author}
  {\bibfnamefont {E.}~\bibnamefont {Thrane}}, \ and\ \bibinfo {author}
  {\bibfnamefont {I.}~\bibnamefont {Mandel}},\ }\href {\doibase
  10.3847/2041-8213/ac2f3c} {\bibfield  {journal} {\bibinfo  {journal} {The
  Astrophysical Journal Letters}\ }\textbf {\bibinfo {volume} {921}},\ \bibinfo
  {pages} {L15} (\bibinfo {year} {2021})}\BibitemShut {NoStop}%
\bibitem [{\citenamefont {Adamcewicz}\ \emph {et~al.}(2025)\citenamefont
  {Adamcewicz}, \citenamefont {Guttman}, \citenamefont {Lasky},\ and\
  \citenamefont {Thrane}}]{adamcewicz2025both}%
  \BibitemOpen
  \bibfield  {author} {\bibinfo {author} {\bibfnamefont {C.}~\bibnamefont
  {Adamcewicz}}, \bibinfo {author} {\bibfnamefont {N.}~\bibnamefont {Guttman}},
  \bibinfo {author} {\bibfnamefont {P.~D.}\ \bibnamefont {Lasky}}, \ and\
  \bibinfo {author} {\bibfnamefont {E.}~\bibnamefont {Thrane}},\ }\href
  {\doibase https://doi.org/10.48550/arXiv.2509.04706} {\bibfield  {journal}
  {\bibinfo  {journal} {arXiv:2509.04706}\ } (\bibinfo {year} {2025}),\
  https://doi.org/10.48550/arXiv.2509.04706}\BibitemShut {NoStop}%
\bibitem [{\citenamefont {Tong}\ \emph {et~al.}(2022)\citenamefont {Tong},
  \citenamefont {Galaudage},\ and\ \citenamefont
  {Thrane}}]{PhysRevD.106.103019}%
  \BibitemOpen
  \bibfield  {author} {\bibinfo {author} {\bibfnamefont {H.}~\bibnamefont
  {Tong}}, \bibinfo {author} {\bibfnamefont {S.}~\bibnamefont {Galaudage}}, \
  and\ \bibinfo {author} {\bibfnamefont {E.}~\bibnamefont {Thrane}},\ }\href
  {\doibase 10.1103/PhysRevD.106.103019} {\bibfield  {journal} {\bibinfo
  {journal} {Phys. Rev. D}\ }\textbf {\bibinfo {volume} {106}},\ \bibinfo
  {pages} {103019} (\bibinfo {year} {2022})}\BibitemShut {NoStop}%
\bibitem [{\citenamefont {Kimball}\ \emph {et~al.}(2020)\citenamefont
  {Kimball}, \citenamefont {Talbot}, \citenamefont {L.~Berry}, \citenamefont
  {Carney}, \citenamefont {Zevin}, \citenamefont {Thrane},\ and\ \citenamefont
  {Kalogera}}]{Kimball_2020}%
  \BibitemOpen
  \bibfield  {author} {\bibinfo {author} {\bibfnamefont {C.}~\bibnamefont
  {Kimball}}, \bibinfo {author} {\bibfnamefont {C.}~\bibnamefont {Talbot}},
  \bibinfo {author} {\bibfnamefont {C.~P.}\ \bibnamefont {L.~Berry}}, \bibinfo
  {author} {\bibfnamefont {M.}~\bibnamefont {Carney}}, \bibinfo {author}
  {\bibfnamefont {M.}~\bibnamefont {Zevin}}, \bibinfo {author} {\bibfnamefont
  {E.}~\bibnamefont {Thrane}}, \ and\ \bibinfo {author} {\bibfnamefont
  {V.}~\bibnamefont {Kalogera}},\ }\href {\doibase 10.3847/1538-4357/aba518}
  {\bibfield  {journal} {\bibinfo  {journal} {The Astrophysical Journal}\
  }\textbf {\bibinfo {volume} {900}},\ \bibinfo {pages} {177} (\bibinfo {year}
  {2020})}\BibitemShut {NoStop}%
\bibitem [{\citenamefont {Miller}\ \emph {et~al.}(2020)\citenamefont {Miller},
  \citenamefont {Callister},\ and\ \citenamefont {Farr}}]{Miller_2020}%
  \BibitemOpen
  \bibfield  {author} {\bibinfo {author} {\bibfnamefont {S.}~\bibnamefont
  {Miller}}, \bibinfo {author} {\bibfnamefont {T.~A.}\ \bibnamefont
  {Callister}}, \ and\ \bibinfo {author} {\bibfnamefont {W.~M.}\ \bibnamefont
  {Farr}},\ }\href {\doibase 10.3847/1538-4357/ab80c0} {\bibfield  {journal}
  {\bibinfo  {journal} {The Astrophysical Journal}\ }\textbf {\bibinfo {volume}
  {895}},\ \bibinfo {pages} {128} (\bibinfo {year} {2020})}\BibitemShut
  {NoStop}%
\bibitem [{\citenamefont {Callister}\ \emph {et~al.}(2022)\citenamefont
  {Callister}, \citenamefont {Miller}, \citenamefont {Chatziioannou},\ and\
  \citenamefont {Farr}}]{Callister_2022}%
  \BibitemOpen
  \bibfield  {author} {\bibinfo {author} {\bibfnamefont {T.~A.}\ \bibnamefont
  {Callister}}, \bibinfo {author} {\bibfnamefont {S.~J.}\ \bibnamefont
  {Miller}}, \bibinfo {author} {\bibfnamefont {K.}~\bibnamefont
  {Chatziioannou}}, \ and\ \bibinfo {author} {\bibfnamefont {W.~M.}\
  \bibnamefont {Farr}},\ }\href {\doibase 10.3847/2041-8213/ac847e} {\bibfield
  {journal} {\bibinfo  {journal} {The Astrophysical Journal Letters}\ }\textbf
  {\bibinfo {volume} {937}},\ \bibinfo {pages} {L13} (\bibinfo {year}
  {2022})}\BibitemShut {NoStop}%
\bibitem [{\citenamefont {Callister}\ and\ \citenamefont
  {Farr}(2024)}]{PhysRevX.14.021005}%
  \BibitemOpen
  \bibfield  {author} {\bibinfo {author} {\bibfnamefont {T.~A.}\ \bibnamefont
  {Callister}}\ and\ \bibinfo {author} {\bibfnamefont {W.~M.}\ \bibnamefont
  {Farr}},\ }\href {\doibase 10.1103/PhysRevX.14.021005} {\bibfield  {journal}
  {\bibinfo  {journal} {Phys. Rev. X}\ }\textbf {\bibinfo {volume} {14}},\
  \bibinfo {pages} {021005} (\bibinfo {year} {2024})}\BibitemShut {NoStop}%
\bibitem [{\citenamefont {Roulet}\ \emph {et~al.}(2021)\citenamefont {Roulet},
  \citenamefont {Chia}, \citenamefont {Olsen}, \citenamefont {Dai},
  \citenamefont {Venumadhav}, \citenamefont {Zackay},\ and\ \citenamefont
  {Zaldarriaga}}]{PhysRevD.104.083010}%
  \BibitemOpen
  \bibfield  {author} {\bibinfo {author} {\bibfnamefont {J.}~\bibnamefont
  {Roulet}}, \bibinfo {author} {\bibfnamefont {H.~S.}\ \bibnamefont {Chia}},
  \bibinfo {author} {\bibfnamefont {S.}~\bibnamefont {Olsen}}, \bibinfo
  {author} {\bibfnamefont {L.}~\bibnamefont {Dai}}, \bibinfo {author}
  {\bibfnamefont {T.}~\bibnamefont {Venumadhav}}, \bibinfo {author}
  {\bibfnamefont {B.}~\bibnamefont {Zackay}}, \ and\ \bibinfo {author}
  {\bibfnamefont {M.}~\bibnamefont {Zaldarriaga}},\ }\href {\doibase
  10.1103/PhysRevD.104.083010} {\bibfield  {journal} {\bibinfo  {journal}
  {Phys. Rev. D}\ }\textbf {\bibinfo {volume} {104}},\ \bibinfo {pages}
  {083010} (\bibinfo {year} {2021})}\BibitemShut {NoStop}%
\bibitem [{\citenamefont {Banagiri}\ \emph
  {et~al.}(2025{\natexlab{b}})\citenamefont {Banagiri}, \citenamefont
  {Callister}, \citenamefont {Adamcewicz}, \citenamefont {Doctor},\ and\
  \citenamefont {Kalogera}}]{Banagiri_2025}%
  \BibitemOpen
  \bibfield  {author} {\bibinfo {author} {\bibfnamefont {S.}~\bibnamefont
  {Banagiri}}, \bibinfo {author} {\bibfnamefont {T.~A.}\ \bibnamefont
  {Callister}}, \bibinfo {author} {\bibfnamefont {C.}~\bibnamefont
  {Adamcewicz}}, \bibinfo {author} {\bibfnamefont {Z.}~\bibnamefont {Doctor}},
  \ and\ \bibinfo {author} {\bibfnamefont {V.}~\bibnamefont {Kalogera}},\
  }\href {\doibase 10.3847/1538-4357/adf4c6} {\bibfield  {journal} {\bibinfo
  {journal} {The Astrophysical Journal}\ }\textbf {\bibinfo {volume} {990}},\
  \bibinfo {pages} {147} (\bibinfo {year} {2025}{\natexlab{b}})}\BibitemShut
  {NoStop}%
\bibitem [{\citenamefont {Sadiq}\ \emph {et~al.}(2025)\citenamefont {Sadiq},
  \citenamefont {Dent},\ and\ \citenamefont
  {Lorenzo-Medina}}]{sadiq2025seeking}%
  \BibitemOpen
  \bibfield  {author} {\bibinfo {author} {\bibfnamefont {J.}~\bibnamefont
  {Sadiq}}, \bibinfo {author} {\bibfnamefont {T.}~\bibnamefont {Dent}}, \ and\
  \bibinfo {author} {\bibfnamefont {A.}~\bibnamefont {Lorenzo-Medina}},\ }\href
  {\doibase https://doi.org/10.48550/arXiv.2506.02250} {\bibfield  {journal}
  {\bibinfo  {journal} {arXiv:2506.02250}\ } (\bibinfo {year} {2025}),\
  https://doi.org/10.48550/arXiv.2506.02250}\BibitemShut {NoStop}%
\bibitem [{\citenamefont {Vitale}\ and\ \citenamefont
  {Mould}(2025)}]{drsl-n3wz}%
  \BibitemOpen
  \bibfield  {author} {\bibinfo {author} {\bibfnamefont {S.}~\bibnamefont
  {Vitale}}\ and\ \bibinfo {author} {\bibfnamefont {M.}~\bibnamefont {Mould}}
  (\bibinfo {collaboration} {Society of Physicists Interested in Non-aligned
  Spins, SPINS}),\ }\href {\doibase 10.1103/drsl-n3wz} {\bibfield  {journal}
  {\bibinfo  {journal} {Phys. Rev. D}\ }\textbf {\bibinfo {volume} {112}},\
  \bibinfo {pages} {083015} (\bibinfo {year} {2025})}\BibitemShut {NoStop}%
\bibitem [{\citenamefont {P\"urrer}\ \emph {et~al.}(2016)\citenamefont
  {P\"urrer}, \citenamefont {Hannam},\ and\ \citenamefont
  {Ohme}}]{PhysRevD.93.084042}%
  \BibitemOpen
  \bibfield  {author} {\bibinfo {author} {\bibfnamefont {M.}~\bibnamefont
  {P\"urrer}}, \bibinfo {author} {\bibfnamefont {M.}~\bibnamefont {Hannam}}, \
  and\ \bibinfo {author} {\bibfnamefont {F.}~\bibnamefont {Ohme}},\ }\href
  {\doibase 10.1103/PhysRevD.93.084042} {\bibfield  {journal} {\bibinfo
  {journal} {Phys. Rev. D}\ }\textbf {\bibinfo {volume} {93}},\ \bibinfo
  {pages} {084042} (\bibinfo {year} {2016})}\BibitemShut {NoStop}%
\bibitem [{\citenamefont {Gupta}(2024)}]{Gupta_2024}%
  \BibitemOpen
  \bibfield  {author} {\bibinfo {author} {\bibfnamefont {I.}~\bibnamefont
  {Gupta}},\ }\href {\doibase 10.3847/1538-4357/ad49a0} {\bibfield  {journal}
  {\bibinfo  {journal} {The Astrophysical Journal}\ }\textbf {\bibinfo {volume}
  {970}},\ \bibinfo {pages} {12} (\bibinfo {year} {2024})}\BibitemShut
  {NoStop}%
\bibitem [{pyc()}]{pycbcsoft}%
  \BibitemOpen
  \href@noop {} {}\bibinfo {howpublished}
  {\url{https://doi.org/10.5281/zenodo.10473621}}\BibitemShut {NoStop}%
\bibitem [{\citenamefont {{LVK Collaboration}}(2018)}]{lalsuite}%
  \BibitemOpen
  \bibfield  {author} {\bibinfo {author} {\bibnamefont {{LVK Collaboration}}},\
  }\href {\doibase 10.7935/GT1W-FZ16} {\enquote {\bibinfo {title} {{LVK}
  {A}lgorithm {L}ibrary - {LALS}uite},}\ }\bibinfo {howpublished} {Free
  software (GPL)} (\bibinfo {year} {2018})\BibitemShut {NoStop}%
\bibitem [{o4a({\natexlab{b}})}]{o4apopzenodo}%
  \BibitemOpen
  \href@noop {} {}\bibinfo {howpublished}
  {\url{https://doi.org/10.5281/zenodo.16053640}} ({\natexlab{b}})\BibitemShut
  {NoStop}%
\bibitem [{\citenamefont {Collaboration}\ \emph
  {et~al.}(2025{\natexlab{c}})\citenamefont {Collaboration}, \citenamefont
  {Collaboration}, \citenamefont {Collaboration} \emph
  {et~al.}}]{ligo2025open}%
  \BibitemOpen
  \bibfield  {author} {\bibinfo {author} {\bibfnamefont {L.~S.}\ \bibnamefont
  {Collaboration}}, \bibinfo {author} {\bibfnamefont {V.}~\bibnamefont
  {Collaboration}}, \bibinfo {author} {\bibfnamefont {K.}~\bibnamefont
  {Collaboration}},  \emph {et~al.},\ }\href {\doibase
  https://doi.org/10.48550/arXiv.2508.18079} {\bibfield  {journal} {\bibinfo
  {journal} {arXiv:2508.18079}\ } (\bibinfo {year} {2025}{\natexlab{c}}),\
  https://doi.org/10.48550/arXiv.2508.18079}\BibitemShut {NoStop}%
\bibitem [{gwo({\natexlab{b}})}]{gwosc}%
  \BibitemOpen
  \href@noop {} {}\bibinfo {howpublished} {\url{https://gwosc.org/}}
  ({\natexlab{b}})\BibitemShut {NoStop}%
\end{thebibliography}%
\bibliographystyle{apsrev4-1}
\end{document}